\documentclass[sort&compress,final,3p,times,twocolumn,fixfloat]{elsarticle}

\usepackage{graphicx}% eps fig, package for more complicated commands
\usepackage{amssymb} % various useful mathematical symbols
\usepackage{amsmath} % extended theorem environments
\usepackage{comment}
\usepackage{lineno}
 \usepackage{ifpdf}
 \ifpdf
 \usepackage[pdftex]{hyperref}
 \else
 \usepackage[hypertex]{hyperref}
 \fi

 \hypersetup{
   pdftitle={},%
   pdfauthor={},%
   pdfsubject={},%
   pdfkeywords={},%
   pdfstartview={},%
   bookmarksopen=true, breaklinks=true, debug=true, %
   colorlinks=true, linkcolor=red, citecolor=blue, urlcolor=blue
 }

\usepackage{float} % improved interface for floating objects
\usepackage[caption=false]{subfig} % supersedes the subfigure package
\newcount\savefnused
\newcount\savefndone

\newcommand{\savefootnote}[2][\empty]% #1=number (optional), #2=text
{\ifx\empty#1\footnotemark\else\footnotemark[#1]\fi
 \global\advance\savefnused by 1
 \expandafter\xdef\csname savefnmark\the\savefnused\endcsname{\thefootnote}%
 \expandafter\xdef\csname savefntext\the\savefnused\endcsname{#2}%
}
\newcommand{\flushfootnote}{\loop\ifnum\savefndone<\savefnused
  \global\advance\savefndone by 1
  \footnotetext[\csname savefnmark\the\savefndone\endcsname]%
    {\csname savefntext\the\savefndone\endcsname}%
  \global\expandafter\let\csname savefnmark\the\savefndone\endcsname\relax
  \global\expandafter\let\csname savefntext\the\savefndone\endcsname\relax
\repeat}

\usepackage{multirow,bigdelim}
\usepackage{tabularx,ragged2e}%

\newcolumntype{Y}{>{\centering\arraybackslash}X}

\def\GeV{{\rm GeV}}
\def\GeV2{{\rm GeV}^2}

\renewcommand{\d}{\mathrm{d}}
\newcommand{\sT}{{\scriptscriptstyle T}}

\def\slash#1{\setbox0=\hbox{$#1$}  % set a box for #1
   \dimen0=\wd0     % and get its size
   \setbox1=\hbox{/} \dimen1=\wd1  % get size of /
   \ifdim\dimen0>\dimen1   % #1 is bigger
      \rlap{\hbox to \dimen0{\hfil/\hfil}} % so center / in box
      #1     % and print #1
   \else     % / is bigger
      \rlap{\hbox to \dimen1{\hfil$#1$\hfil}} % so center #1
      /      % and print /
   \fi}      %  
   
\usepackage[normalem]{ulem}

\newcommand{\new}[1]{{\color{blue}#1}}

\begin{document}

\begin{frontmatter}

\title{The transverse nucleon single-spin asymmetry for the semi-inclusive production of photons in lepton-nucleon scattering}

\author[a,b]{Weeam S. Albaltan}
\ead{wbaltan@nmsu.edu}
\address[a]{Department of Physics, New Mexico State University, Las Cruces, NM 88003-0001, USA}
\address[b]{Department of Physics, College of Sciences, Princess Nourah bint Abdulrahman University (PNU), Riyadh 11671, Saudi Arabia}
\author[c,d]{Alexei Prokudin}
\ead{prokudin@jlab.org}
\address[c]{Division of Science, Penn State University Berks, Reading, Pennsylvania 19610, USA}
\address[d]{Theory Center, Jefferson Lab, 12000 Jefferson Avenue, Newport News, Virginia 23606, USA}
\author[a]{Marc Schlegel}
\ead{schlegel@nmsu.edu}

%\begin{linenumbers}

\begin{abstract}\small
{We study the semi-inclusive production of real, high-$p_T$, isolated photons in unpolarized and polarized lepton-proton collisions, $\ell p\to \ell\gamma X$. In particular we analyze the transverse nucleon single-spin asymmetry within the collinear twist-3 formalism in perturbative QCD  to leading order (LO) accuracy. We find that this spin asymmetry is generated by twist-3 dynamical quark-gluon-quark ($qgq$) correlations in the nucleon through the so-called soft-fermion and hard pole contributions. Hence, this process unprecedentedly allows for a point-by-point scan of the support of the dynamical $qgq$ twist-3 matrix elements $F_{FT}(x,x^\prime)$ and $G_{FT}(x,x^\prime)$ in lepton-nucleon scattering experiments. 
}
\end{abstract}

\begin{keyword}
Electron-Ion Collider,
transverse single-spin asymmetries,
lepton-nucleon scattering,
gamma SIDIS,
twist-3,
JLAB-THY-19-3067
\end{keyword}

\end{frontmatter}

%%%%%%%%%%%%%%%%%%%%%%%%%%%%%%%%%%%%%%%%%%%%%%%%%%%%%%%%%%%%%%%%%%%%%%%%%%%%
\section{Introduction}

Transverse single-spin asymmetries (SSA) in single-inclusive high-energy collisions have received a lot of attention in recent years. In particular, remarkably large SSAs have been reported in polarized proton collisions, $pp^{\uparrow}\to h X$, where $p^\uparrow$ denotes a transversely polarized proton (see review \cite{Aidala:2012mv}). The theoretical treatment within perturbative QCD (pQCD) $\--$ based on the so-called collinear twist-3 formalism $\--$ is challenging for processes like $pp^{\uparrow}\to h X$ that are mediated purely by the strong force, see Refs.~ \cite{Qiu:1991pp,Qiu:1991wg,Kanazawa:2000hz,Kouvaris:2006zy,Kang:2010zzb,Kanazawa:2011bg,Beppu:2010qn,Metz:2012ct,Yuan:2009dw,Kanazawa:2013uia,Kanazawa:2014dca}. The main reason for this is that several competing matrix elements describing multiparton correlations in each colliding and produced hadron enter such factorization formulae. Thus, it is not an easy task to extract information on these multiparton correlation functions from proton-proton data alone.\\
A reduction of complexity of (underlying) processes may be achieved by studying transverse spin effects in high-energy lepton-nucleon collisions. A promising experimental facility to do so would be the future Electron-Ion Collider (EIC) \cite{Accardi:2012qut}. Naively, at first sight, the easiest transverse single-spin observable is an SSA in inclusive deep-inelastic lepton ($\ell$) - nucleon ($N$) scattering (DIS), $\ell N^\uparrow \to \ell X$. However, it is a peculiarity of this observable that time-reversal symmetry forces the SSA to vanish in Quantum Electrodynamics (QED) for a one-photon exchange between lepton and nucleon \cite{Christ:1966zz}. A non-zero transverse nucleon SSA in DIS may be generated by higher-order QED contributions due to a two-photon exchange \cite{Metz:2006pe,Afanasev:2007ii,Metz:2012ui,Schlegel:2012ve}, however the additional photon exchange causes a suppression factor for the SSA of one order of the QED fine structure constant, $\alpha_{\mathrm{em}}\simeq 1/137$. A naturally small effect of about $10^{-4} - 10^{-3}$ is expected for the proton SSA \cite{Schlegel:2014qza}. Surprisingly, somewhat larger effects have been found for a neutron target at JLab \cite{Katich:2013atq}, while a small effect was indeed observed for a proton target at HERMES \cite{:2009wj}.\\
Another promising reaction that may give access to multiparton correlation functions at an EIC is the (polarized) single-inclusive production of hadrons or jets from leptons scattering off protons, $\ell N \to hX$ \cite{Gamberg:2014eia,Kanazawa:2014tda,Kanazawa:2015jxa,Kanazawa:2015ajw}. In particular the polarized production of jets in this process might provide valuable information on soft-gluon pole contributions at LO in pQCD. It is however expected that NLO and NNLO corrections can become sizable for jet production at an EIC \cite{Hinderer:2015hra,Abelof:2016pby}, and an NLO calculation for the transverse nucleon SSA has not yet been presented in the literature. On the other hand, if a hadron instead of a jet is produced, multiparton correlations in the fragmentation process also contribute to the asymmetry \cite{Gamberg:2014eia,Kanazawa:2015ajw}, and it might turn out that the fragmentation contribution may potentially be the dominant source for a measured SSA, see Ref.~\cite{Kanazawa:2014dca}. At NLO, Ref.~\cite{Hinderer:2017ntk} questions how well we understand the origin of polarized processes in inclusive hadron production for leptons scattering off nucleons.\\
In this paper we investigate the semi-inclusive production process of real, high-$p_T$, isolated photons in lepton-nucleon collisions which we refer to in the following as $\gamma$SIDIS. The $\ell p\to \ell\gamma X$ process and major sources of its background were considered previously in the literature, see Refs.~\cite{Brodsky:1972yx,deRujula:1998yq,Hoyer:2000mb,Mukherjee:2004dw}. $\gamma$SIDIS combines two positive features of the aforementioned processes: 
\begin{enumerate}
\item Due to the additional radiation of a photon, a transerse nucleon SSA in this process will not suffer the same fate as the SSA in DIS, that is, to vanish under time-reversal symmetry at LO. Consequently, the SSA will not be suppressed by $\alpha_{\mathrm{em}}$. 
\item  Since we consider a real, high-$p_T$, isolated photon in the final state emitted in a point-like QED-vertex rather than in a fragmentation process, the only non-perturbative objects generating the transverse nucleon SSA will be quark-gluon-quark correlation functions in the nucleon. Hence, we expect that  these functions can be cleanly probed in the (polarized) semi-inclusive production process of photons.
\end{enumerate}

Our paper is organized as follows: In Section \ref{sec:II} we analytically calculate the unpolarized cross section to LO accuracy and set up the kinematics and notations. In Section \ref{sec:III} we present our analytical LO results for the transverse nucleon-spin dependent cross section calculated in the collinear twist-3 formalism. Finally, in Section \ref{sec:conclusion} we present our conclusions.

\section{Analytical LO formula for the unpolarized cross section\label{sec:II}}
%%%%%%%%%%%%%%%%%%%%%%%%%%%%%%%%%%%%%%%%%%%%%%%%%%%%%%%%%%%%%%%%%%%%%%%%%%%%
\begin{figure}
\centering
\includegraphics[width=0.5\columnwidth]{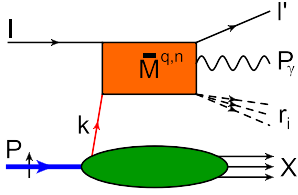}
\vspace*{-0cm}
\caption{Factorized amplitude with one quark leg connecting the nucleon (green blob) and the hard scattering amplitude $\bar{\mathcal{M}}^{q,n}$.
}
\label{fig:qqGen}
\end{figure}
%%%%%%%%%%%%%%%%%%%%%%%%%%%%%%%%%%%%%%%%%%%%%%%%%%%%%%%%%%%%%%%%%%%%%%%%%%%
 In this section we perform a LO calculation of the unpolarized cross section for the process $\ell (l)+N(P)\to \ell(l^\prime)+\gamma(P_\gamma)+X$. As indicated, the four momenta of the incident lepton and nucleon are labeled as $l^\mu$ and $P^\mu$, respectively. In the final state we assume that the lepton is detected with a four momentum $l^{\prime\mu}$ along with a real, high-$p_T$, isolated photon with four momentum $P_\gamma^\mu$. Since we are interested in this process at high energies, we will neglect the lepton and nucleon mass and treat their four momenta as light-like vectors, $l^2=l^{\prime2}=P^2\simeq 0$. Of course, we also have $P_\gamma^2=0$ for a real photon.

We apply the common factorization procedure to the amplitude of this process, sketched in Fig.~\ref{fig:qqGen}. Here, the full amplitude is factorized into a hard scattering amplitude $\bar{\mathcal{M}}^q$ which is subject to perturbation theory, and a hadronic matrix element describing the soft physics in the nucleon. Along with the detected lepton and photon we take into account that there may be additional undetected partons produced in the hard scattering whose momenta  $\lbrace r_i \rbrace$ are integrated out.\\
By squaring the amplitude in Fig.~\ref{fig:qqGen} we can derive a formula for the fully differential cross section. We may write this factorized cross section as an integral over the full four momentum $k^\mu$ of a hard partonic cross section $\hat{\sigma}^{\bar{q}q}(k)$ and a quark-quark correlator $\Phi(k)$. The partonic cross section may be defined as
\begin{eqnarray}
\hspace{-0.4cm}\hat{\sigma}^{\bar{q}q}_{rs}(k) & = & \sum_{n=1}^\infty \prod_{i=1}^n\int \frac{d^4 r_i}{(2\pi)^3}\,\Theta(r^0_i)\,\delta(r_i^2-m_i^2)\,\times\label{eq:partCSqqDef}\\
&&\delta^{(4)}(k+q-R)\,\left[\bar{\mathcal{M}}^{q,n}_s(k)\,(\gamma^0 \bar{\mathcal{M}}^{q,n}(k))^\dagger_r\right]\,.\nonumber
\end{eqnarray}
In this formula we encounter the partonic amplitude  $\bar{\mathcal{M}}^{q,n}$ which we consider the sum of all Feynman diagrams for a subprocess $\ell q\to\ell \gamma +\mathrm{n\,unobserved\,partons}$. The quark legs in these Feynman diagrams are amputated, that is why the amplitudes carry an open Dirac index $r$ or $s$. The integrals in Eq.~\eqref{eq:partCSqqDef} refer to the Lorentz-invariant phase space integrations over the parton momenta $r_1,...,r_n$. The produced unobserved partons may or may not be massive (mass $m_i$). The sum of the unobserved parton momenta is labeled by $R=\sum_{i=1}^nr_i$, while the momentum $q$ is defined as $q=l-l^\prime-P_\gamma$.

We then apply the {\it collinear approximation} to the quark momentum $k^\mu$ in terms of a Sudakov decomposition, 
\begin{eqnarray}
k^{\mu}\simeq x\,P^{\mu}+k_\sT^{\mu}\,.\label{eq:collexpk}
\end{eqnarray}
This decomposition provides the collinear picture where the parton in a nucleon moves collinearly to the nucleon with a momentum fraction $x$. In addition, small {\it transverse} (to the nucleon's motion) deviations from this picture are allowed through the (small) transverse momentum $k_\sT$ of the parton. In order to define the term {\it transverse} an additional light-cone vector $n^\mu$ is required satisfying $n^2=0$, and for a normalization we choose $P\cdot n=1$. Then, {\it transverse} is defined through the projector
\begin{eqnarray}
g_\sT^{\mu\nu}=g^{\mu\nu}-P^\mu n^\nu-P^\nu n^\mu\,.\label{eq:trProj}
\end{eqnarray}
Any transverse vector is defined by $a_\sT^\mu=g_\sT^{\mu\nu}a_\nu$. A component of $k$ in the direction of $n$ is neglected in Eq.~\eqref{eq:collexpk}. In fact, for twist-2 observables like the unpolarized cross section it would be sufficient to consider $k\simeq xP$.

Once the collinear approximation $k\simeq xP$ is applied to the partonic cross section (\ref{eq:partCSqqDef}) we can write the factorized unpolarized cross section as
\begin{eqnarray}
\hspace{-0.5cm}\mathrm{D}\sigma=\frac{1}{32\pi^2\,s}\int\d x\,\hat{\sigma}_{rs}^{\bar{q}q}(k=xP)\,\Phi_{sr}^q(x)+\mathcal{O}(\Lambda/Q)\,,\label{eq:CStw2}
\end{eqnarray}
where we introduced the Mandelstam variable $s=(P+l)^2$ as well as the short notation $\mathrm{D}\sigma=E^\prime E_\gamma\frac{d\sigma}{d^3\vec{l}^\prime\,d^3\vec{P}_\gamma}$. $E^\prime$ and $E_\gamma$ are the energies of the detected lepton and photon, respectively. Furthermore, the hard scale of the process is $Q^2=-q^2$, and $\Lambda$ is some small hadronic scale, for example the nucleon mass $M$. We also encounter the unpolarized collinear correlator \cite{Kanazawa:2015ajw},
\begin{eqnarray}
\hspace{-0.5cm}\Phi^q_{sr}(x)&=&\int_{-\infty}^\infty\frac{d\lambda}{2\pi}\,\mathrm{e}^{i\lambda x}\langle P|\,\bar{q}_r(0)\,[0;\lambda n]\,q_s(\lambda n)\,|P\rangle\nonumber\\
&=&\tfrac{1}{2}\slash{P}_{sr}\,f_1^q(x).\label{eq:PDF}
\end{eqnarray}
In this definition $q,\bar{q}$ denote quark fields and $[0;\lambda n]$ a Wilson line along the light-cone direction $n$. The function $f_1^q(x)$ is the usual unpolarized parton distribution function (PDF) for a quark flavor $q$.

%%%%%%%%%%%%%%%%%%%%%%%%%%%%%%%%%%%%%%%%%%%%%%%%%%%%%%%%%%%%%%%%%%%%%%%%%%%%
\begin{figure}
\centering
\includegraphics[width=0.5\columnwidth]{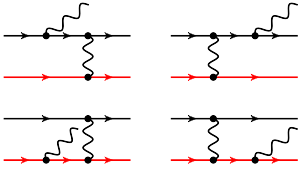}
\vspace*{-0cm}
\caption{LO diagrams for the subprocess $\ell q\to \ell q \gamma$. The photon can either be radiated of the lepton (black fermion line) or the quark (red fermion line).
}
\label{fig:qLO}
\end{figure}
%%%%%%%%%%%%%%%%%%%%%%%%%%%%%%%%%%%%%%%%%%%%%%%%%%%%%%%%%%%%%%%%%%%%%%%%%%%

We evaluate the partonic cross section to LO accuracy. The corresponding diagrams are shown in Fig. \ref{fig:qLO}. This calculation has already been performed early on in the seventies by Brodsky, Gunion and Jaffe (see Ref.~\cite{Brodsky:1972yx}). We repeated this calculation as a check, and fully confirm their results. In order to compare our results to their calculation we use the following Lorentz-invariant dimensionless variables introduced in \cite{Brodsky:1972yx},
\begin{eqnarray}
P\cdot l=\tfrac{1}{2}Q^2\,\alpha&;&P\cdot l^\prime=\tfrac{1}{2}Q^2\,\alpha^\prime\,\nonumber\\
l\cdot P_\gamma=\tfrac{1}{2}Q^2\,\beta&;&l^\prime\cdot P_\gamma=\tfrac{1}{2}Q^2\,\beta^\prime\,\nonumber\\
P\cdot P_\gamma = \tfrac{1}{2}Q^2\,\gamma &;& \tilde{Q}^2=(1-\beta+\beta^\prime)\,Q^2\,.\label{eq:BrodVar}
\end{eqnarray}
Another four vector is introduced, $\tilde{q}^\mu=(l-l^\prime)^\mu$, with $\tilde{Q}^2=-\tilde{q}^2$ being a different scale that would correspond to the hard scale in a DIS process. It was discussed in detail in Ref.~\cite{Brodsky:1972yx} that a parton model picture emerges if both scales $Q^2$ and $\tilde{Q}^2$ are larger than the nucleon mass $M$, as well as their difference,
\begin{eqnarray}
Q^2\gg M^2\,;&\tilde{Q}^2\gg M^2\,;&Q^2-\tilde{Q}^2\gg M^2\,.\label{eq:hardscales} 
\end{eqnarray}

As discussed in Ref.~\cite{Brodsky:1972yx}, every partonic factor that appears in factorization formulae for the semi-inclusive photon production process receives three kinds of contributions: 
\begin{enumerate}
\item The square of diagrams where the photon is radiated off the lepton $\--$ Bethe-Heitler (BH) contributions, 
\item The square of diagrams where the photon is radiated off the quark $\--$ Compton (C) contributions, 
\item Interfering diagrams where the photon is radiated off the lepton and quark $\--$ Interference (I) contributions.
\end{enumerate}
Those three effects are cleanly separated by the fractional quark charges. While the quark charges appear squared for the BH contribution, $e_q^2$, their powers are larger for C- and I-contributions, i.e., $e_q^4$ and $e_q^3$, respectively. For this reason we define the following charge combinations of a generic correlation function $f^{q}$ of quark flavor $q$ according to the charge weightings:
\begin{eqnarray}
f^{\mathrm{BH}}&\equiv&\sum_{q=u,d,s,...}e_q^2\,\left(f^{q}+f^{\bar{q}}\right)\,,\nonumber\\
f^{\mathrm{C}}&\equiv&\sum_{q=u,d,s,...}e_q^4\,\left(f^{q}+f^{\bar{q}}\right)\,,\nonumber\\
f^{\mathrm{I}}&\equiv&\sum_{q=u,d,s,...}e_q^3\,\left(f^{q}-f^{\bar{q}}\right)\,.\label{eq:chargePDFs}
\end{eqnarray} 
Note that there is a relative sign between quark and anti-quark functions in the flavor combination $f^{\mathrm{I}}$. It is not, a priori, obvious at LO at which factorization scale $\mu$ the correlation functions are to be evaluated in Eq.~\eqref{eq:chargePDFs}. While a choice $\mu=Q$ and $\mu=\tilde{Q}$ seems to be plausible for the BH- and C-contributions it is not so clear for the interference contributions I. Hence, we take the geometric average $\mu=\sqrt{Q\tilde{Q}}$ as a plausible choice. 

At LO the fully differential unpolarized cross section has the following form,
\begin{eqnarray}
\mathrm{D}\sigma_{\mathrm{U}}^\mathrm{LO}=\frac{\alpha^3_{\mathrm{em}}}{4\pi^2\,s\,Q^4}\sum_{k=\mathrm{BH,C,I}}\hat{\sigma}^k_{\mathrm{U}}\,f_1^k(x_B)\,.\label{eq:UnpCS}
\end{eqnarray}
The parton distributions $f_1^k$ that appear in Eq.~\eqref{eq:UnpCS} are evaluated at a specific {\it scaling variable} $x_B$ which we define as
\begin{eqnarray}
x_B=\frac{Q^2}{2P\cdot q}=\frac{-(l-l^\prime-P_\gamma)^2}{2P\cdot (l-l^\prime-P_\gamma)}=\frac{1}{\alpha-\alpha^\prime-\gamma}\,.\label{eq:xB}
\end{eqnarray}
The value of $x_B$ is entirely driven by the external kinematics and one can, just like in DIS, scan the PDF point-by-point. Note however that the scaling variable $x_B$ is different from the one we encounter in DIS,
\begin{eqnarray}
\tilde{x}_B \equiv \frac{\tilde{Q}^2}{2P\cdot \tilde{q}}=\frac{-(l-l^\prime)^2}{2P\cdot(l-l^\prime)}=\frac{1-\beta+\beta^\prime}{\alpha-\alpha^\prime}\,,\label{eq:xBt}
\end{eqnarray}
 due to the additional photon emission.

The partonic cross sections $\hat{\sigma}^k_{\mathrm{U}}$ depend on the external kinematical variables defined in Eq.~\eqref{eq:BrodVar}. They were presented in Ref.~\cite{Brodsky:1972yx}. For the convenience of the reader we show the explicit expressions in \ref{sec:appendix}.

\section{Analytical LO formula for the twist-3 transverse nucleon SSA\label{sec:III}}
%%%%%%%%%%%%%%%%%%%%%%%%%%%%%%%%%%%%%%%%%%%%%%%%%%%%%%%%%%%%%%%%%%%%%%%%%%%%
\begin{figure}
\centering
\includegraphics[width=0.5\columnwidth]{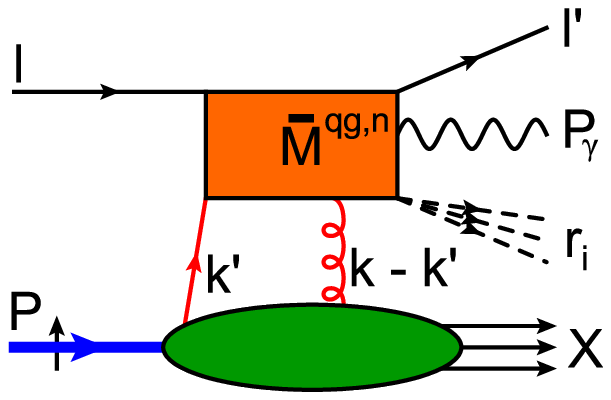}
\vspace*{-0cm}
\caption{Factorized amplitude with a quark and a gluon leg connecting the nucleon (green blob) and the hard scattering amplitude $\bar{\mathcal{M}}^{qg,n}$.
}
\label{fig:qgqGen}
\end{figure}
%%%%%%%%%%%%%%%%%%%%%%%%%%%%%%%%%%%%%%%%%%%%%%%%%%%%%%%%%%%%%%%%%%%%%%%%%%%

In this section we present our result for the transverse nucleon SSA  calculated within the collinear twist-3 formalism. In general, a twist-3 observable may receive contributions from more than just one class of correlation functions. Typically, there are three classes (cf. Ref.~\cite{Kanazawa:2015ajw}): {\it intrinsic}, {\it kinematical} and {\it dynamical} twist-3 contributions. For the particular observable we are interested in it is easy to see that there are no intrinsic twist-3 contributions. However, kinematical twist-3 contributions may very well contribute. They are generated by a Taylor expansion ({\it collinear expansion}) of the partonic cross section in terms of the transverse parton momentum $k_\sT$ in Eq.~(\ref{eq:collexpk}) up to first order (twist-3),
\begin{eqnarray}
\hspace{-0.6cm}\hat{\sigma}^{\bar{q}q}_{rs}(k)=\hat{\sigma}^{\bar{q}q}_{rs}(xP)+k_\sT^{\rho}\frac{\partial \hat{\sigma}^{\bar{q}q}_{rs}(xP+k_\sT)}{\partial k_\sT^\rho}\Bigg|_{k_\sT=0}+\mathcal{O}(k_\sT^2)\,.
\end{eqnarray}
The zeroth order generates the intrinsic twist-3 contribution, and as stated above, it vanishes for an unpolarized incident lepton and transversely polarized nucleon. For the first order, at LO the derivative w.r.t. $k_\sT$ may also act on a delta function in Eq.~(\ref{eq:partCSqqDef}). Performing algebraic manipulations, this derivative may be turned into an $x$ - derivative on the kinematical twist-3 matrix element. We find the following form of the kinematical twist-3 contribution,
\begin{eqnarray}
\mathrm{D}\sigma^{\mathrm{LO}}_{\mathrm{T,kin}}&=&\frac{\alpha^3_{\mathrm{em}}}{4\pi^2\,s\,Q^4}\,M\,\epsilon^{Pn\rho S}\label{eq:kintw3}\\
&&\hspace{-1.7cm}\,\sum_{k=\mathrm{BH,C,I}}\Bigg[\left(\frac{\partial\,\hat{\sigma}^k_{\mathrm{U}}}{\partial k_\sT^\rho}-\frac{2\,x_B\,q_{\sT \rho}}{Q^2}\frac{\partial\,\hat{\sigma}^k_{\mathrm{U}}}{\partial x} \right)^{x=x_B}_{k_\sT=0}\,f_{1T}^{\perp (1), k}(x_B)\nonumber\\
&&-\hat{\sigma}^k_{\mathrm{U}}\Big|^{x=x_B}_{k_\sT=0}\,\frac{2x_B \,q_{\sT \rho}}{Q^2}\frac{d}{dx_B}f_{1T}^{\perp (1), k}(x_B)\Bigg]\,, \nonumber
\end{eqnarray}
where  the totally antisymmetric tensor $\epsilon^{Pn\rho S}=\epsilon^{\mu\nu\rho\sigma}P_\mu n_\nu S _\sigma$ with the sign convention $\epsilon^{0123}=+1$ is introduced. The vector $S^\mu$ denotes the four dimensional nucleon spin vector with $P\cdot S=0$ and $S^2=-1$.

In Eq.~\eqref{eq:kintw3} the partonic functions $\hat{\sigma}^{k}_{\mathrm{U}}$ are calculated in the same way as the ones in Eq.~\eqref{eq:UnpCS} for the unpolarized cross section, however the momentum $k$ that appears in Eq.~\eqref{eq:partCSqqDef} is not immediately set to $x_B P$ but to $xP+k_\sT$, {\it then} the derivatives in Eq.~\eqref{eq:kintw3} are performed, and {\it afterwards} we set $x=x_B$ and $k_\sT=0$.

The correlation function $f_{1T}^{\perp (1)}(x)$ is the first $k_\sT$-moment of the well-known Sivers function \cite{Sivers:1989cc,Sivers:1990fh}. It is defined as follows,
\begin{eqnarray}
\Phi^{\rho,q}_{\partial,sr}(x)&=&\int d^2 k_\sT\,k_\sT^\rho\,\Phi^q_{sr}(x,k_\sT^2)\label{eq:partialPhi}\\
&=&\tfrac{1}{2}\,M\,\epsilon^{Pn\rho S}\,\slash P _{sr}\,f_{1T}^{\perp (1),q}(x)+...\,.\nonumber
\end{eqnarray}
The (naive) definition of the transverse momentum dependent (TMD) quark-quark correlator $\Phi(x,k_\sT^2)$ in Eq.~\eqref{eq:partialPhi} including a future-pointing Wilson line (needed for semi-inclusive DIS), as well as its parameterization, may be found in Refs.~\cite{Goeke:2005hb,Bacchetta:2006tn}. 

There are also dynamical twist-3 contributions that are generated by quark-gluon-quark correlations. Those contributions are interferences of amplitudes shown in Fig.~\ref{fig:qgqGen} (with a quark {\it and} a gluon entering the hard scattering) and in Fig.~\ref{fig:qqGen} (with only one quark entering the hard scattering part). The quark and the gluon in Fig.~\ref{fig:qgqGen} carry momenta $k^\prime$ and $k-k^\prime$, respectively. We may apply the collinear approximation like Eq.~\eqref{eq:collexpk} to these momenta right away, so that $k^\mu\simeq xP^\mu$ and $k^{\prime \mu}\simeq x^\prime P^\mu$. A convenient gauge for a gluon field $A^\mu$ to work with in the collinear twist-3 formalism is the light-cone gauge $n\cdot A=0$. Using this procedure, we derive a general factorization formula like Eq.~\eqref{eq:CStw2}, but for the dynamical twist-3 contribution based on Fig.~\ref{fig:qgqGen}:
\begin{eqnarray}
\mathrm{D}\sigma_{\mathrm{T,dyn}}&=&\frac{-1}{32\pi^2 s}\int \d x\int\d x^\prime\,\label{eq:CStw3dyn}\\&&
\hspace{-1.5cm}\left(\hat{\sigma}_{rs,\rho}^{\bar{q}gq}(k=xP,k^\prime=x^\prime P)\,\frac{i\,\Phi_{F,sr}^{q,\rho}(x,x^\prime)}{x^\prime-x}+\mathrm{c.c.}\right)\,.\nonumber
\end{eqnarray}
The partonic cross section is similarly defined as Eq.~\eqref{eq:partCSqqDef}, but with a hard quark-gluon scattering amplitude,
\begin{eqnarray}
\hspace{-0.4cm}\hat{\sigma}^{\bar{q}gq}_{rs,\rho}(k,k^\prime) & = & \sum_{n=1}^\infty \prod_{i=1}^n\int \frac{d^4 r_i}{(2\pi)^3}\,\Theta(r_i^0)\,\delta(r_i^2-m_i^2)\,\times\nonumber\\
&&\hspace{-1.6cm}\delta^{(4)}(k+q-R)\,\left[\bar{\mathcal{M}}^{qg,n}_{s,\rho}(k,k^\prime)\,(\gamma^0 \bar{\mathcal{M}}^{q,n}(k))^\dagger_r\right]\,.\label{eq:partCSqgqDef}
\end{eqnarray}
The quark-gluon-quark correlator $\Phi_F(x,x^\prime)$ appears in Eq.~\eqref{eq:CStw3dyn} and its definition and parameterization can be found in \cite{Kanazawa:2015ajw}. Here we just write the relevant correlation functions for a transversely polarized nucleon,
\begin{eqnarray}
\Phi^{q,\rho}_{rs,F}(x,x^\prime)&=&\tfrac{1}{2}\,M\,i\,\epsilon^{Pn\rho S}\,\slash{P}_{rs}\,F_{FT}^q(x,x^\prime)\label{eq:PhiF}\\
&&-\tfrac{1}{2}\,M\,S_\sT^{\rho}\,(\slash{P}\gamma_5)_{rs}\,G_{FT}^q(x,x^\prime)+...\,.\nonumber
\end{eqnarray}

%%%%%%%%%%%%%%%%%%%%%%%%%%%%%%%%%%%%%%%%%%%%%%%%%%%%%%%%%%%%%%%%%%%%%%%%%%%%
\begin{figure}
\centering
\includegraphics[width=\columnwidth]{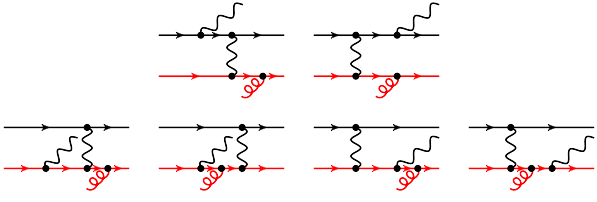}
\vspace*{-0cm}
\caption{LO diagrams for the subprocess $\ell q g\to \ell q \gamma$. The hard photon can either be radiated of the lepton (black fermion line) or the quark (red fermion line).
}
\label{fig:qgLO}
\end{figure}
%%%%%%%%%%%%%%%%%%%%%%%%%%%%%%%%%%%%%%%%%%%%%%%%%%%%%%%%%%%%%%%%%%%%%%%%%%%
In Fig.~\ref{fig:qgLO} we show the LO diagrams relevant for the quark-gluon hard amplitude $\bar{\mathcal{M}}^{qg,1}_{s,\rho}$ in Eq.~\eqref{eq:partCSqgqDef} where they are to be interfered with the diagrams of Fig.~\ref{fig:qLO}. At LO ($n=1$), just like the unpolarized case, the partonic cross sections (\ref{eq:partCSqgqDef}) will be proportional to a delta function $\delta(x-x_B)$, fixing the momentum fraction $x=x_B$. In addition, we note that all numerators of the interference cross section $\hat{\sigma}^{\bar{q}gq}$ are either imaginary (when being traced with $\slash{P}$) or real (when being traced with $\slash{P}\gamma_5$). Since we add the complex conjugate ($\mathrm{c.c.}$) in Eq.~\eqref{eq:CStw3dyn} an additional imaginary part needs to be generated to obtain a non-vanishing transverse spin dependence. This additional imaginary part is induced by the $i\epsilon$ part that appears in  the quark propagators in Fig.~\ref{fig:qgLO}, where we encounter three different types of propagators,
\begin{eqnarray}
\hspace{-0.5cm}\frac{1}{(k^\prime+q)^2+i\epsilon}&=&\frac{x_B}{Q^2}\left(\frac{\mathcal{P}}{x^\prime - x_B}-i\pi \delta(x^\prime - x_B)\right)\,,\nonumber\\
\hspace{-0.5cm}\frac{1}{(k^\prime+\tilde{q})^2+i\epsilon}&=&\frac{\tilde{x}_B}{\tilde{Q}^2}\left(\frac{\mathcal{P}}{x^\prime-\tilde{x}_B}-i\pi\delta(x^\prime-\tilde{x}_B)\right)\,,\nonumber\\
\hspace{-0.5cm}\frac{1}{(k^\prime-P_\gamma)^2+i\epsilon}&=&-\frac{1}{\gamma Q^2}\left(\frac{\mathcal{P}}{x^\prime}+i\pi\delta(x^\prime)\right)\,.\label{eq:poles}
\end{eqnarray}
The delta-function in the first line of Eq.~\eqref{eq:poles} produces the so-called {\it soft-gluonic pole} (SGP) contributions, the second line generates {\it hard-pole} (HP) contributions, while the third line generates {\it soft-fermionic pole} (SFP) contributions.

The SGP contributions fix the second momentum fraction in $\Phi_{F,sr}^{q,\rho}(x,x^\prime)$ to be $x^\prime=x_B$, and those contributions will be proportional to the $qgq$ correlation function $F_{FT}(x_B,x_B)$ ({\it Efremov-Teryaev-Qiu-Sterman} (ETQS) matrix element \cite{Efremov:1981sh,Qiu:1991pp,Qiu:1991wg})\footnote{Note that no contribution proportional to $G_{FT}(x_B,x_B)$ appears since this function vanishes due to the anti-symmetry property of that particular function, $G_{FT}(x,x^\prime)=-G_{FT}(x^\prime,x)$. In contrast, we have $F_{FT}(x,x^\prime)=+F_{FT}(x^\prime,x)$ \cite{Kanazawa:2015ajw}.}. Also, the derivative terms $\frac{d}{dx_B}F_{FT}(x_B,x_B)$ may appear through terms proportional to $\delta(x^\prime-x_B)/(x^\prime-x_B)=-\frac{d}{dx^\prime}\delta(x^\prime - x_B)$, and a subsequent integration by parts in $x^\prime$. The SGP contributions and the kinematical twist-3 contributions in \new{Eq.~}\eqref{eq:kintw3} can be added using the well-known connection between the Sivers function and the ETQS-matrix element \cite{Boer:2003cm,Kanazawa:2015ajw,Gamberg:2017jha},
\begin{eqnarray}
\pi\,F_{FT}^q(x,x)=f_{1T}^{\perp (1),q}(x)\,.\label{eq:RelSivQS}
\end{eqnarray}
We find that those two terms cancel each other, i.e., the SGP and kinematical twist-3 contribution from Eq.~\eqref{eq:kintw3}, for all three channels $BH$, $C$ and $I$ when being summed. As a consequence, SGP terms do not enter the transverse spin dependent cross section, and the HP and SFP terms are the only remaining contributions. 

In general, the light-cone vector $n^\mu$ explicitly enters partonic twist-3 cross sections through the $\epsilon^{Pn\rho S}$ and $S_T^\rho$ terms in the parameterization (\ref{eq:PhiF}). We checked the independence on the choice of the light-cone vector $n^\mu$ of the partonic HP and SFP cross sections. To do so we followed the procedure of Ref.~\cite{Kanazawa:2015ajw} and wrote the $n^\mu$ vector as a linear combination of the four linearly independent physical vectors $P$, $l$, $l^\prime$, $P_\gamma$,
\begin{eqnarray}
n^\mu = a\,l^\mu+b\,P^\mu+c\,l^{\prime \mu}+d\,P_\gamma^\mu\,.\label{eq:n}
\end{eqnarray} 
The conditions $P\cdot n=1$ and $n^2=0$ eliminate only two of the coefficients, say, $a$ and $b$ in favor of $c$ and $d$, which cannot be otherwise determined. One might worry that the coefficients $c$ and $d$ enter the partonic cross sections in this way, which would lead to the explicit dependence on vector $n$, and hence to ambiguities. We checked explicitly that this is not the case and that the coefficients $c$ and $d$ do cancel out. The remaining parts of the vector $n$ in Eq.~\eqref{eq:n} that enter the cross sections amount to a choice $n^\mu=\frac{2}{\alpha Q^2}l^\mu$. That would be the natural choice in a lepton-nucleon center-of-mass frame.

After adding all twist-3 contributions the final LO formula for the spin-dependent cross section emerges,
\begin{eqnarray}
\hspace{-0.6cm}\mathrm{D} \sigma_{\mathrm{T}}^{\mathrm{LO}}(S)&=&\frac{\alpha_{\mathrm{em}}^3}{4\pi^2\,s\,Q^4}\,\frac{\pi M}{Q}\,\Big[\frac{\epsilon^{Pll^\prime S}}{Q^3}\,\sigma_{\mathrm{UT}}^1+\frac{\epsilon^{PlP_{\gamma} S}}{Q^3}\,\sigma_{\mathrm{UT}}^2\nonumber\\
&&\hspace{-1.5cm}+\frac{(l_\sT^\prime\cdot S_\sT)\epsilon^{Pll^\prime P_{\gamma}}}{Q^5}\,\sigma_{\mathrm{UT}}^3+\frac{(P_{\gamma \sT}\cdot S_\sT)\epsilon^{Pll^\prime P_{\gamma}}}{Q^5}\,\sigma_{\mathrm{UT}}^4\Big].\label{eq:CSTSF}
\end{eqnarray}
Even though Eq.~\eqref{eq:CSTSF} is written completely covariantly, the structures $\sigma_{\mathrm{UT}}^i$ may be viewed as different azimuthal modulations of the transverse spin vector $S_\sT^\mu$. At this point we present the analytical expression for those modulations,
\begin{eqnarray}
\hspace{-0.6cm}\sigma_{\mathrm{UT}}^{i=1,2}&=&\sum_{k=\mathrm{C,I}}\Big[\hat{\sigma}_{\mathrm{HP,F}}^{i,k}\,F_{FT}^k(x_B,\tilde{x}_B)+\hat{\sigma}_{\mathrm{SFP,F}}^{i,k}\,F_{FT}^k(x_B,0)\nonumber\\
&&\hspace{-0.6cm}+\hat{\sigma}_{\mathrm{HP,G}}^{i,k}\,G_{FT}^k(x_B,\tilde{x}_B)+\hat{\sigma}_{\mathrm{SFP,G}}^{i,k}\,G_{FT}^k(x_B,0)\Big]\,,\label{eq:CSUT12}\\
\hspace{-0.6cm}\sigma_{\mathrm{UT}}^{i=3,4}&=&\sum_{k=\mathrm{C,I}}\Big[\hat{\sigma}_{\mathrm{HP,G}}^{i,k}\,G_{FT}^k(x_B,\tilde{x}_B)\nonumber\\
&&\hspace{2.4cm}+\hat{\sigma}_{\mathrm{SFP,G}}^{i,k}\,G_{FT}^k(x_B,0)\Big].\label{eq:CSUT34}
\end{eqnarray}
We note that only Compton (C) and Interference (I) effects contribute to the transverse nucleon spin dependent cross section Eq.~\eqref{eq:CSTSF}. The Bethe-Heitler (BH) contributions to $\mathrm{D}\sigma_{\mathrm{UT}}$ where the detected photon is radiated off the leptonic part cancel once dynamical- and kinematical twist-3 contributions are added. This feature is in agreement with the situation for inclusive DIS where a transverse single-spin asymmetry is forbidden for a one-photon exchange due to time-reversal symmetry \cite{Christ:1966zz}.  The first and second lines of Eq.~\eqref{eq:CSUT12} display HP and SFP contributions of the $qgq$ correlation functions $F_{FT}$ and $G_{FT}$, respectively. The structures $\sigma_{\mathrm{UT}}^{i=3,4}$ are generated by HP and SFP contributions of the $qgq$ correlation functions $G_{FT}$ alone. The partonic factors $\hat{\sigma}^{i,k}$ depend on external kinematical variables (\ref{eq:BrodVar}). Their analytical expressions are rather lengthy, and we present our results in \ref{sec:appendix}.

Eqs.~(\ref{eq:CSTSF},\ref{eq:CSUT12},\ref{eq:CSUT34}) constitute the main result of this paper. They reveal the possibility to experimentally scan the $qgq$ correlation functions $F_{FT}$ and $G_{FT}$ point-by-point at LO on their full support in $x$ and $x'$, by varying the scaling variable $x_B$ and $\tilde{x}_B$. This means that a measurement of the transverse nucleon SSA, e.g., at an EIC, can give us for the first time direct information on these functions. Such a feature is unprecedented and does not appear in other processes in this way, to the best of our knowledge.

As the variables defined in Eq.~\eqref{eq:BrodVar} are not intuitive, we want to rewrite the cross sections Eq.~\eqref{eq:UnpCS} and Eq.~\eqref{eq:CSTSF} in terms of the experimentally accessible kinematical variables
that typically appear in collider experiment like an EIC, i.e., transverse momentum $p_\sT^{\prime,\gamma}$, pseudo-rapidity $\eta^{\prime,\gamma}$ and azimuthal angle $\phi^{\prime,\gamma}$ of the scattered electron and the isolated photon. The momenta and nucleon spin vector acquire the following form in the lepton-nucleon center-of-mass (cm) frame,
\begin{eqnarray}
\hspace{-1cm}&P^\mu=\tfrac{1}{2}\sqrt{s}\left(1,0,0,1\right)\,,\,l^\mu=\tfrac{1}{2}\sqrt{s}\left(1,0,0,-1\right)\,,&\nonumber\\
\hspace{-1cm}&l^{\prime \mu}=p_\sT^\prime\left(\cosh \eta^{\prime},\cos\phi^\prime,\sin\phi^\prime,\sinh\eta^\prime\right)\,,&\nonumber\\
&P_\gamma^\mu=p_\sT^\gamma\left(\cosh \eta^{\gamma},\cos\phi^\gamma,\sin\phi^\gamma,\sinh\eta^\gamma\right)\,,&\nonumber\\
\hspace{-1cm}&S^\mu=\left(0,\cos\phi_s,\sin\phi_s,0\right)\,.&\label{eq:cmVector}
\end{eqnarray} 
The Lorentz-invariant cross sections $\mathrm{D}\sigma$ in Eqs.~(\ref{eq:UnpCS},\ref{eq:CSTSF}) can be translated to the cm-frame as follows,
\begin{eqnarray}
\frac{\d\sigma}{\d p_\sT^\prime\,\d\eta^\prime\,\d\phi^\prime\,\d p_\sT^\gamma\,\d\eta^\gamma\,\d\phi^\gamma}=p_\sT^\prime\,p_\sT^\gamma\,\mathrm{D}\sigma\,,\label{eq:cmCS}
\end{eqnarray}
where $\mathrm{D}\sigma$ depends on the variables defined in Eq.~\eqref{eq:BrodVar}. Of course, those variables need to be expressed explicitly in terms of the cm-variables Eq.~\eqref{eq:cmVector}. We note that the unpolarized cross section,  Eq.~\eqref{eq:UnpCS}, depends only on the difference of the azimuthal angles $\phi^\prime-\phi^\gamma$, and one may integrate out one of those angles. On the other hand, the spin-dependent cross section, Eq.~\eqref{eq:CSTSF}, carries four azimuthal structures related to the transverse spin vector $S$. While the functions $\sigma_{\mathrm{UT}}^i$ in Eq.~\eqref{eq:CSTSF} only depend on the difference of the azimuthal angles $\phi^\prime-\phi^\gamma$ $\--$ like the unpolarized cross section $\--$ the prefactors include the azimuthal angle $\phi_s$ of the nucleon spin vector in the following way,
\begin{eqnarray}
\hspace{-0.5cm}\frac{\epsilon^{Pll^\prime S}}{Q^3}&=&\tfrac{1}{2}\sqrt{\alpha\alpha^\prime(1-\beta+\beta^\prime)}\sin(\phi_s-\phi^\prime)\,,\nonumber\\
\hspace{-0.5cm}\frac{\epsilon^{PlP_\gamma S}}{Q^3}&=&\tfrac{1}{2}\sqrt{\alpha\beta\gamma}\sin(\phi_s-\phi^\gamma)\,,\nonumber\\
\hspace{-0.5cm}\frac{(l^\prime_\sT\cdot S_\sT)\epsilon^{Pll^\prime P_\gamma}}{Q^5}&=&\tfrac{1}{2}\tfrac{\alpha^\prime}{\alpha}(1-\beta+\beta^\prime)\sqrt{\alpha\beta\gamma}\times\nonumber\\
&&\sin(\phi^\prime-\phi^\gamma)\cos(\phi_s-\phi^\prime)\,,\nonumber\\
\hspace{-0.5cm}\frac{(P_{\gamma\sT}\cdot S_\sT)\epsilon^{Pll^\prime P_\gamma}}{Q^5}&=&\tfrac{1}{2}\sqrt{\alpha\alpha^\prime(1-\beta+\beta^\prime)}\tfrac{\beta\gamma}{\alpha}\times\nonumber\\
&&\sin(\phi^\prime-\phi^\gamma)\cos(\phi_s-\phi^\gamma)\,.\label{eq:cmphi}
\end{eqnarray}
By varying the azimuthal angle $\phi_s$ it will be possible to experimentally disentangle the four azimuthal structures.

\section{Conclusions\label{sec:conclusion}}

In this paper we have calculated the transverse nucleon spin-dependent cross section of the semi-inclusive production process of real, high-$p_\sT$, isolated photons in lepton-nucleon collisions, $\ell N^\uparrow\to\ell \gamma X$, to LO accuracy in the collinear twist-3 formalism. We found that a non-zero transverse spin-dependent cross section is generated by the quark-gluon-quark correlation functions $F_{FT}^q$ and $G_{FT}^q$. Most importantly, through measurements (e.g., at the EIC) of the momentum spectra of leptons and photons it might be possible to (partially) reconstruct these otherwise unknown functions point-by-point from experimental data. In this sense the suggested process plays the same important role for the determination of the $qgq$ functions as DIS has historically played (and still does) for the determination of quark PDFs. We also stress out that information on the support of the $qgq$ functions $F_{FT}^q$ and $G_{FT}^q$ will have a great impact on the evolution of TMD functions like the Sivers function \cite{Braun:2009mi, Vogelsang:2009pj,Kang:2010xv, Aybat:2011ge} and will deepen our understanding on the $g_2$ structure function in DIS.

Since the functions $F_{FT}^q(x,x^\prime)$ and $G_{FT}^q(x,x^\prime)$ are essentially unknown for $x\neq x^\prime$ we leave the
estimate of the transverse nucleon SSA at an EIC for a future publication. We expect that the total cross section of $\gamma$SIDIS will be suppressed with respect to DIS cross section in the same (lepton) bin due to
 a suppression factor of $\alpha_{\mathrm{em}}$ for $\gamma$SIDIS compared to DIS. However, since the expected event DIS rate is large at an EIC we conclude that $\gamma$SIDIS may be feasible as well. 

We are also aware that the LO calculation considered in this paper is likely not sufficient for a detailed description of data gathered at the EIC. In particular, it is known that gluons (at NLO) play an important role \cite{Hoyer:2000mb}. At NLO, other (unknown) matrix elements, such as triple-gluon correlations might be probed when measuring the nucleon SSA in  $\gamma$SIDIS. Also, the isolated photon may be produced exclusively through a soft quark-antiquark distribution amplitude at NLO, rather than in a point-like QED vertex. However, such effects are beyond the scope of this paper, and we leave the study of those NLO effects for a future publication.

%%%%%%%%%%%%%%%%%%%%%%%%%%%%%%%%%%%%%%%%%%%%%%%%%%%%%%%%%%%%%%%%%%%%%%%%%%%%

\section*{Acknowledgements}
The authors would like to acknowledge useful discussions with Andreas Metz and Charles Hyde.
A.P. is supported by the National Science Foundation
under Grant No.~PHY-1623454 and the DOE Contract No. DE-
AC05-06OR23177, under which Jefferson Science Associates, LLC operates
Jefferson Lab.
This work was also supported by the U.S. Department of Energy through the TMD Topical Collaboration.

%\end{linenumbers}

\appendix

\section{The full expressions for the hard partonic factors} \label{sec:appendix}
In this appendix we present our analytical LO results for the various partonic cross sections discussed in the main text.
\subsection{Unpolarized cross section}
We abbreviate the denominators $D_1=1-\beta+\beta^\prime$, $D_2=1-x_B(\alpha-\alpha^\prime)$ and $D_3=1-\beta+\beta^\prime-x_B(\alpha-\alpha^\prime)$, and obtain
\begin{eqnarray}
\hspace{-0.6cm}\hat{\sigma}_{\mathrm{U}}^{\mathrm{BH}}&=&\frac{4}{\beta  \beta^{\prime}} \Big(2(1-\beta)+\beta ^2 +\beta^{\prime 2}+2 \beta^{\prime}\label{eq:partCSUBH}\\
&&-2 x_B
   (\alpha  (1+ \beta^{\prime})-\alpha^{\prime} (1-\beta))\nonumber\\
&&+2 x_B^2 \left(\alpha ^2+\alpha^{\prime 2}\right)\Big)\,,\nonumber\\
\hspace{-0.6cm}\hat{\sigma}_{\mathrm{U}}^{\mathrm{C}}&=&\frac{\beta \beta^\prime\,\hat{\sigma}_{\mathrm{U}}^{\mathrm{BH}}}{D_1\,D_2\,D_3} \,,\label{eq:partCSUC}\\
\hspace{-0.6cm}\hat{\sigma}_\mathrm{U}^\mathrm{I}&=&-\frac{a_U+b_U\,x_B+c_U\,x_B^2+d_U x_B^3}{\beta\,\beta^\prime \,D_1\,D_2\,D_3}\,,\label{eq:partCSUI}
\end{eqnarray}
with the following coefficients,
\begin{eqnarray}
\hspace{-0.6cm}a_U&=&4\,D_1\,(\beta+\beta^\prime)(2-\beta(2-\beta)+\beta^\prime(2+\beta^\prime))\,,\label{eq:aU}
\end{eqnarray}
\begin{eqnarray}
\hspace{-0.6cm}b_U&=& 4\alpha\,\Big[(\beta -1) \beta  ((\beta -2) \beta +4)\nonumber\\
&&-\,\beta^\prime  (\beta -2) ((\beta -3) \beta -2)+\beta^{\prime 2}((\beta -3) \beta -6)\nonumber\\
&&-\,\beta^{\prime 3}(\beta +3)\Big]+4\alpha^\prime\,\Big[ \beta  (3 (\beta -2) \beta +4)\nonumber\\
&&-\beta^\prime  (\beta -2) ((\beta -1) \beta +2)+ \beta^{\prime 2}(\beta -3) (\beta -2)\nonumber\\
&&- \beta^{\prime 3}(\beta -3)+\beta^{\prime 4}\Big]\,,\label{eq:bU}
\end{eqnarray}
\begin{eqnarray}
\hspace{-0.6cm}c_U&=&8\alpha^2\Big[2 (1-\beta) \beta+\beta^\prime (2+\beta(2-\beta) )+\beta^{\prime 2}(2+\beta)\Big]\nonumber\\
&&\hspace{-0.3cm}+8\alpha^{\prime 2}\Big[2 (1-\beta) \beta+\beta^\prime (2-\beta(2-\beta))+\beta^{\prime 2}(2-\beta)\Big]\nonumber\\
&&\hspace{-0.3cm}+8\alpha\,\alpha^\prime\Big[-\beta(2-\beta(2-\beta))-\beta^\prime(2-\beta^2)\nonumber\\
&&\hspace{1cm}-\beta^{\prime 2}(2-\beta)-\beta^{\prime 3}\Big]\,,\label{eq:cU}
\end{eqnarray}
\begin{eqnarray}
\hspace{-0.6cm}d_U&=&8(\alpha^2+\alpha^{\prime 2})\Big[\alpha^\prime (\beta+\beta^\prime(1-\beta)+\beta^{\prime 2})\nonumber\\
&&-\alpha(\beta(1-\beta)+\beta^\prime\,(1+\beta))\Big]\,,\label{eq:dU}
\end{eqnarray}

\subsection{Transversely polarized cross section - Compton contributions}  
Here we give our explicit analytical results for the  partonic factors of the Compton contributions in Eqs. (\ref{eq:CSUT12},\ref{eq:CSUT34}). The ones appearing in the structure $\sigma^1_{\mathrm{UT}}$ read,
\begin{eqnarray}
\hspace{-0.7cm}\hat{\sigma}^{\mathrm{1,C}}_{\mathrm{HP,F}}&=& \frac{16}{\alpha\,D_1^2\,(1-D_2)\,D_3^2}\,\Big[-D_1^2-2x_B\,\alpha^\prime(1+\beta^\prime)D_1\nonumber\\
&&\hspace{-0.6cm} +2x_B^2\,\left(\alpha^2\beta(1-\beta)+\alpha\alpha^\prime(1+2\beta^\prime)-\alpha^{\prime 2}(1+\beta^\prime)^2\right)\nonumber\\
&&\hspace{-0.6cm}-x_B^3\,D_1(\alpha-\alpha^\prime)(\alpha^2+\alpha^{\prime 2})\Big],\label{eq:sig1CHPF}\\
\hspace{-0.7cm}\hat{\sigma}^{\mathrm{1,C}}_{\mathrm{SFP,F}}&=& -\frac{16}{\alpha\,D_1^2\,D_2\,D_3}\,\Big[\nonumber\\
&&\hspace{0cm}-D_1(3-2\beta(1-\beta)+2\beta^\prime(2+\beta^\prime))\nonumber\\
&&\hspace{0cm}+x_B\Big(D_1(3\alpha D_1-\alpha^\prime(2+D_1))\nonumber\\
&&\hspace{0cm}-2\alpha\beta(1-2D_1)+2\beta^2(\alpha-\alpha^\prime)\Big)\nonumber\\
&&-x_B^2\,D_1(\alpha^2+\alpha^{\prime 2})\Big],\label{eq:sig1CSFPF}
\end{eqnarray}
\begin{eqnarray}
\hspace{-0.7cm}\hat{\sigma}^{\mathrm{1,C}}_{\mathrm{HP,G}}&=& \frac{16}{\alpha\,D_1\,(1-D_2)\,D_3^2}\,\Big[D_1+2x_B\alpha(1+\beta^\prime)\nonumber\\
&&-2x_B^2\alpha\left((\alpha-2\alpha^\prime)(1-\beta)-\alpha^\prime\beta^\prime\right)\nonumber\\
&&-x_B^3\,(\alpha-\alpha^\prime)(\alpha^2+\alpha^{\prime 2})\Big],\label{eq:sig1CHPG}\\
\hspace{-0.7cm}\hat{\sigma}^{\mathrm{1,C}}_{\mathrm{SFP,G}}&=& -\frac{16}{\alpha\,D_1\,D_2\,D_3}\,\Big[-(1+2D_1)\nonumber\\
&&-x_B\left(\alpha^\prime(2+D_1)-\alpha(3-3\beta+\beta^\prime)\right)\nonumber\\
&&-x_B^2\,(\alpha^2+\alpha^{\prime 2})\Big].\label{eq:sig1CSFPG}
\end{eqnarray}
The partonic functions for the structure $\sigma^2_{\mathrm{UT}}$ read
\begin{eqnarray}
\hspace{-0.7cm}\hat{\sigma}^{\mathrm{2,C}}_{\mathrm{HP,F}}&=& \frac{16}{\alpha\,(\alpha-\alpha^\prime)\,D_1^2\,D_2\,D_3^2}\,\Big[D_1^2(\alpha-\alpha^\prime)\nonumber\\
&&-2x_B\,D_1\left(\alpha^2(1-\beta)+\alpha^{\prime 2}(1+\beta^\prime)\right)\nonumber\\
&&+2x_B^2\,(\alpha-\alpha^\prime)\,\Big(\alpha^2(1-\beta)^2+\alpha^{\prime 2}(1+\beta^\prime)^2\nonumber\\
&&+\alpha\alpha^\prime(1+\beta^\prime-\beta(1+2\beta^\prime))\Big)\nonumber\\
&&+x_B^3\,D_1(\alpha-\alpha^{\prime})^2(\alpha^2+\alpha^{\prime 2})\Big],\label{eq:sig2CHPF}\\
\hspace{-0.7cm}\hat{\sigma}^{\mathrm{2,C}}_{\mathrm{SFP,F}}&=& -\frac{16x_B}{\alpha\,D_1^2\,D_2^2\,D_3}\,\Big[\nonumber\\
&&D_1\Big(\alpha(3-2\beta(2-\beta)+2\beta^\prime(1+\beta^\prime))\nonumber\\
&&-\alpha^\prime(3-2\beta(1-\beta)+2\beta^\prime(2+\beta^\prime))\Big)\nonumber\\
&&-x_B\Big(\alpha^2\left(3+\beta^2-2\beta(2+\beta^\prime)+\beta^\prime(4+3\beta^\prime)\right)\nonumber\\
&&-2\alpha\alpha^\prime\left(3-\beta(5-2\beta)+5\beta^\prime-2\beta\beta^\prime+2\beta^{\prime 2}\right)\nonumber\\
&&+\alpha^{\prime 2}\left(D_1(2+D_1)+2\beta^2\right)\Big)\nonumber\\
&&+x_B^2\,D_1(\alpha-\alpha^\prime)(\alpha^2+\alpha^{\prime 2})\Big],\label{eq:sig2CSFPF}
\end{eqnarray}
\begin{eqnarray}
\hspace{-0.7cm}\hat{\sigma}^{\mathrm{2,C}}_{\mathrm{HP,G}}&=& \frac{16x_B}{\alpha\,D_1\,D_2\,(1-D_2)\,D_3^2}\,\Big[-D_1(\alpha-\alpha^\prime)\nonumber\\
&&+2x_B\,\alpha\alpha^\prime (1+D_1)\nonumber\\
&&-2x_B^2\,\alpha\alpha^\prime(\alpha-\alpha^\prime)(1+2D_1)\nonumber\\
&&+x_B^3\,(\alpha-\alpha^\prime)^2(\alpha^2+\alpha^{\prime 2})\Big],\label{eq:sig2CHPG}\\
\hspace{-0.7cm}\hat{\sigma}^{\mathrm{2,C}}_{\mathrm{SFP,G}}&=& -\frac{16x_B}{\alpha\,D_1\,D_2^2\,D_3}\,\Big[(1+2D_1)(\alpha-\alpha^\prime)\nonumber\\
&&-x_B\Big((\alpha^2+\alpha^{\prime 2})(2+D_1)-2\alpha\alpha^\prime (1+2D_1)\Big)\nonumber\\
&&+x_B^2\,(\alpha-\alpha^\prime)(\alpha^2+\alpha^{\prime 2})\Big].\label{eq:sig2CSFPG}
\end{eqnarray}
For the structure $\sigma^3_{\mathrm{UT}}$ we obtain the following partonic factors,
\begin{eqnarray}
\hspace{-0.7cm}\hat{\sigma}^{\mathrm{3,C}}_{\mathrm{HP,G}}&=& -\frac{64x_B(1+\beta^\prime)}{\,D_1^2\,D_3^2}\,,\label{eq:sig3CHPG}\\
\hspace{-0.7cm}\hat{\sigma}^{\mathrm{3,C}}_{\mathrm{SFP,G}}&=&  -\frac{64x_B(1+\beta^\prime)}{\,D_1^2\,D_2\,D_3}\,,\label{eq:sig3CSFPG}
\end{eqnarray}
while for the structure $\sigma^4_{\mathrm{UT}}$ we obtain
\begin{eqnarray}
\hspace{-0.7cm}\hat{\sigma}^{\mathrm{4,C}}_{\mathrm{HP,G}}&=& -\frac{64x_B^2(\alpha+\alpha^\prime-\alpha\beta+\alpha^\prime\beta^\prime)}{\,D_1^2\,D_2\,D_3^2}\,,\label{eq:sig4CHPG}\\
\hspace{-0.7cm}\hat{\sigma}^{\mathrm{4,C}}_{\mathrm{SFP,G}}&=&  -\frac{64x_B^2(\alpha+\alpha^\prime-\alpha\beta+\alpha^\prime\beta^\prime)}{\,D_1^2\,D_2^2\,D_3}\,.\label{eq:sig4CSFPG}
\end{eqnarray}

\subsection{Transversely polarized cross section - Interference contributions}

At last we give our explicit analytical results for the  partonic factors of the Interference contributions in Eqs. (\ref{eq:CSUT12},\ref{eq:CSUT34}). The ones appearing in the structure $\sigma^1_{\mathrm{UT}}$ read,
\begin{eqnarray}
\hspace{-0.7cm}\hat{\sigma}^{\mathrm{1,I}}_{\mathrm{HP,F}}&=& \frac{16}{\alpha\,\beta\,\beta^\prime\,D_1\,(1-D_2)\,D_3^2}\,\Big[\beta^\prime\,D_1^2\nonumber\\
&&\hspace{-0.5cm}+x_B\,D_1 \left(\alpha(\beta^2-(1+\beta)\beta^\prime)+\alpha^\prime\beta^\prime(3-\beta+3\beta^\prime)\right)\nonumber\\
&&\hspace{-0.5cm}-2x_B^2\Big(\alpha^2\beta^2(1-\beta)+\alpha\alpha^\prime\beta^\prime(1+\beta+\beta^\prime+2\beta\beta^\prime)\nonumber\\
&&\hspace{-0.5cm}-\alpha^{\prime 2}\beta^\prime(D_1+\beta)(2D_1+\beta)\Big)\nonumber\\
&&\hspace{-0.5cm}+x_B^3\Big(-\alpha^2\alpha^\prime\beta^\prime(1-\beta-6\beta^2+\beta^\prime)\nonumber\\
&&\hspace{-0.5cm}+\alpha^3(2\beta^2(1-\beta)+\beta^\prime-\beta\beta^\prime+\beta^{\prime 2})\nonumber\\
&&\hspace{-0.5cm}+\alpha^{\prime 3}\beta^\prime(\beta+(1+\beta^\prime)(1+2\beta^\prime))\nonumber\\
&&\hspace{-0.5cm}-\alpha\alpha^{\prime 2}\beta^\prime (1+\beta^\prime+\beta(5+6\beta^\prime))\Big)\nonumber\\
&&\hspace{-0.5cm}+x_B^4(\alpha-\alpha^\prime)(\alpha^2+\alpha^{\prime 2})\Big(\alpha(\beta^2-(1+\beta)\beta^\prime)\nonumber\\
&&\hspace{-0.5cm}+\alpha^\prime \beta^\prime\,D_1\Big)\Big],\label{eq:sig1IHPF}
\end{eqnarray}
\begin{eqnarray}
\hspace{-0.7cm}\hat{\sigma}^{\mathrm{1,I}}_{\mathrm{SFP,F}}&=& -\frac{16}{\alpha\,\beta\,\beta^\prime\,D_1\,D_2}\,\Big[\beta^3-\beta^2(2-\beta^\prime)+\beta^\prime(1+\beta^\prime)^2\nonumber\\
&&\hspace{-0.5cm}+\beta(2+\beta^\prime(2+\beta^\prime))\nonumber\\
&&\hspace{-0.5cm}+x_B\Big(\alpha^\prime(2\beta-\beta^2(2-\beta^\prime)+\beta^\prime(1+\beta^\prime)^2)\nonumber\\
&&\hspace{-0.5cm}+\alpha(\beta^2-\beta^3-\beta^\prime-\beta^{\prime 2}-\beta(2+\beta^\prime(2+\beta^\prime)))\Big)\nonumber\\
&&\hspace{-0.5cm}+x_B^2\,\beta(\alpha^2+\alpha^{\prime 2})\Big],\label{eq:sig1ISFPF}
\end{eqnarray}
\begin{eqnarray}
\hspace{-0.7cm}\hat{\sigma}^{\mathrm{1,I}}_{\mathrm{HP,G}}&=& \frac{16}{\alpha\,\beta\,\beta^\prime\,D_1\,(1-D_2)\,D_3}\,\Big[-\beta^\prime D_1\label{eq:sig1IHPG}\\
&&\hspace{-0.5cm}-x_B\,\Big(\alpha (\beta^{\prime 2}+\beta(2+\beta^\prime))+\alpha^\prime(2D_1+\beta)(D_1-1)\Big)\nonumber\\
&&\hspace{-0.5cm}-x_B^2\,\Big(\alpha^2(\beta^{\prime 2}-\beta(2-\beta^\prime))+\alpha\alpha^\prime(\beta^{\prime 2}+\beta(4-3\beta))\nonumber\\
&&\hspace{-0.5cm}+\alpha^{\prime 2}(2D_1+\beta)(D_1-1)\Big)\nonumber\\
&&\hspace{-0.5cm}-x_B^3\,(\alpha^2+\alpha^{\prime 2})(\alpha(1-D_1-\beta D_1)+\alpha^\prime\beta^\prime D_1)\Big],\nonumber\\
\hspace{-0.7cm}\hat{\sigma}^{\mathrm{1,I}}_{\mathrm{SFP,G}}&=& -\frac{16}{\alpha\,\beta\,\beta^\prime D_1\,D_2}\,\Big[\beta+D_1(\beta+\beta^\prime)\nonumber\\
&&\hspace{-0.5cm}+x_B\Big(\alpha^\prime (\beta+D_1(\beta+\beta^\prime))-\alpha(2\beta(1-\beta)+\beta^\prime)\Big)\nonumber\\
&&\hspace{-0.5cm}+x_B^2(\alpha^2+\alpha^{\prime 2})\beta\Big].\label{eq:sig1ISFPG}
\end{eqnarray}
The partonic factor appearing in the structure $\sigma^2_{\mathrm{UT}}$ read,
\begin{eqnarray}
\hspace{-0.7cm}\hat{\sigma}^{\mathrm{2,I}}_{\mathrm{HP,F}}&=& \frac{16x_B}{\alpha\,\beta\,\beta^\prime\,D_1\,(1-D_2)\,D_3^2}\,\Big[D_1^2(\alpha^\prime\beta^\prime-\alpha\beta)\nonumber\\
&&\hspace{-0.5cm+2x_B\,D_1\Big(\alpha^2\beta(1-\beta)+\alpha^{\prime 2}\beta^\prime(1+\beta^\prime)\Big)}\nonumber\\
&&\hspace{-0.5cm}-2x_B^2\Big(\alpha^3\beta(1-\beta)^2+\alpha^2\alpha^\prime\beta\beta^\prime(2-3\beta)\nonumber\\
&&\hspace{-0.5cm}-\alpha^{\prime 3}\beta^\prime(1+\beta^\prime)^2+\alpha\alpha^{\prime 2}\beta\beta^\prime(2+3\beta^\prime)\Big)\nonumber\\
&&\hspace{-0.5cm}+x_B^3(\alpha-\alpha^\prime)(\alpha^2+\alpha^{\prime 2})(\alpha^\prime\beta^\prime-\alpha\beta)D_1\Big],\label{eq:sig2IHPF}\\
\hspace{-0.7cm}\hat{\sigma}^{\mathrm{2,I}}_{\mathrm{SFP,F}}&=& -\frac{16x_B}{\alpha\,\beta\,\beta^\prime\,D_1\,D_2^2}\,\Big[\alpha^\prime\Big(\beta^3-\beta^2(2-\beta^\prime)\nonumber\\
&&\hspace{-0.5cm}+\beta^\prime(1+\beta^\prime)^2+\beta(2+\beta^\prime(2+\beta^\prime))\Big)\nonumber\\
&&\hspace{-0.5cm}-\alpha\Big(\beta^3-\beta^2(2-\beta^\prime)+\beta(1-\beta^\prime)^2\nonumber\\
&&\hspace{-0.5cm}+\beta^\prime(2+\beta^\prime(2+\beta^\prime))\Big)\nonumber\\
&&\hspace{-0.5cm}+x_B\,\Big(\alpha^2(\beta(1-\beta)^2+2\beta^{\prime}+(2+\beta)\beta^{\prime 2})\nonumber\\
&&\hspace{-0.5cm}+\alpha^{\prime 2}(2\beta-\beta^2(2-\beta^\prime)+\beta^\prime(1+\beta^\prime)^2)\nonumber\\
&&\hspace{-0.5cm}-\alpha\alpha^\prime(\beta+\beta^\prime)(3-\beta(2-\beta)+\beta^\prime(2+\beta^\prime))\Big)\nonumber\\
&&\hspace{-0.5cm}+x_B^2\,(\alpha^2+\alpha^{\prime 2})(\alpha^\prime\beta-\alpha\beta^\prime)\Big],\label{eq:sig2ISFPF}
\end{eqnarray}
\begin{eqnarray}
\hspace{-0.7cm}\hat{\sigma}^{\mathrm{2,I}}_{\mathrm{HP,G}}&=& \frac{16x_B}{\alpha\,\beta\,\beta^\prime\,D_1\,(1-D_2)\,D_3}\,\Big[D_1(\alpha\beta-\alpha^\prime\beta^\prime)\nonumber\\
&&\hspace{-0.5cm}+x_B\Big(\alpha^{\prime 2}(D_1-D_1^2+\beta)-\alpha^2(1-\beta-D_1^2)\nonumber\\
&&\hspace{-0.5cm}-\alpha\alpha^\prime(\beta+\beta^\prime)(3-2\beta+2\beta^\prime)\Big)\nonumber\\
&&\hspace{-0.5cm}+x_B^2(\alpha^2+\alpha^{\prime 2})D_1(\alpha\beta-\alpha^\prime\beta^\prime)\Big],\label{eq:sig2IHPG}\\
\hspace{-0.7cm}\hat{\sigma}^{\mathrm{2,I}}_{\mathrm{SFP,G}}&=& -\frac{16x_B}{\alpha\,\beta\,\beta^\prime D_1\,D_2^2}\,\Big[\alpha^\prime(\beta+D_1(\beta+\beta^\prime))\nonumber\\
&&\hspace{-0.5cm}+\alpha(1-\beta-D_1(1+\beta+\beta^\prime))\nonumber\\
&&\hspace{-0.5cm}-x_B(\alpha-\alpha^\prime)\Big(\alpha^\prime(\beta+D_1(\beta+\beta^\prime))\nonumber\\
&&\hspace{-0.5cm}+\alpha(1-\beta-D_1(1+\beta+\beta^\prime))\Big)\nonumber\\
&&\hspace{-0.5cm}+x_B^2(\alpha^2+\alpha^{\prime 2})(\alpha^\prime\beta-\alpha\beta^\prime)\Big].\label{eq:sig2ISFPG}
\end{eqnarray}
For the structure $\sigma^3_{\mathrm{UT}}$ we obtain the following partonic factors,
\begin{eqnarray}
\hspace{-0.7cm}\hat{\sigma}^{\mathrm{3,I}}_{\mathrm{HP,G}}&=& \frac{32x_B}{\beta\,\beta^\prime\,D_1\,D_3^2}\Big[-\beta^2+\beta^{\prime 2}+2\beta(1+\beta^\prime)\nonumber\\
&&\hspace{-0.5cm}+x_B\Big(\beta(\alpha^\prime(2-\beta)-\alpha(2+\beta))+\beta^{\prime 2}(\alpha+\alpha^\prime)\nonumber\\
&&\hspace{-0.5cm}-2(\alpha-\alpha^\prime)\beta\beta^\prime\Big)\Big]\,,\label{eq:sig3IHPG}\\
\hspace{-0.7cm}\hat{\sigma}^{\mathrm{3,I}}_{\mathrm{SFP,G}}&=&  \frac{32x_B(\beta^2+\beta^\prime(2+\beta^\prime))}{\beta\,\beta^\prime\,D_1\,D_2}\,,\label{eq:sig3ISFPG}
\end{eqnarray}
while for the structure $\sigma^4_{\mathrm{UT}}$ we obtain
\begin{eqnarray}
\hspace{-0.7cm}\hat{\sigma}^{\mathrm{4,I}}_{\mathrm{HP,G}}&=& \frac{32x_B^2}{\beta\,\beta^\prime\,D_1\,D_3^2}\,\Big[\beta(\alpha^\prime(2-\beta)-\alpha\beta)\nonumber\\
&&\hspace{-0.5cm}+2\beta^\prime (\alpha(1-\beta)+\alpha^\prime\beta)+(\alpha+\alpha^\prime)\beta^{\prime 2}\Big],\label{eq:sig4IHPG}\\
\hspace{-0.7cm}\hat{\sigma}^{\mathrm{4,I}}_{\mathrm{SFP,G}}&=&  -\frac{32x_B^2}{\beta\,\beta^\prime\,D_1\,D_2^2}\,\Big[\alpha(\beta^{\prime 2}-\beta(2-\beta))\nonumber\\
&&\hspace{-0.5cm}-\alpha^\prime(\beta^2+\beta^\prime(2+\beta^\prime))\Big].\label{eq:sig4ISFPG}
\end{eqnarray}
\bibliographystyle{Science}

\bibliography{Referenzen}

\clearpage

\end{document}